\documentclass[reprint, amsmath, amssymb,superscriptaddress]{revtex4-1}

\usepackage{soul}

\usepackage{bm}
\usepackage[retainorgcmds]{IEEEtrantools}
\usepackage{graphicx}
\usepackage{float}
\usepackage{mathrsfs}
\usepackage{amsmath}
\usepackage{amssymb}
\usepackage{color}
\usepackage{nicefrac}
\usepackage{lineno}
\usepackage[abs]{overpic}
\usepackage{xcolor,varwidth}
\usepackage[outline]{contour}\contourlength{0.05em}

\newcommand{\ud}{\,\mathrm{d}}

\DeclareMathOperator{\tr}{tr}

\begin{document}

\title{The role of quantum coherence in non-equilibrium entropy production}
\date{\today}
\author{Jader P. Santos}
\affiliation{Instituto de F\'isica da Universidade de S\~ao Paulo,  05314-970 S\~ao Paulo, Brazil}
\author{Lucas C. C\'eleri}
\affiliation{Instituto de F\'isica, Universidade Federal de Goi\'as, Caixa Postal 131, 74001-970, Goi\^ania, Brazil}
\author{Gabriel T. Landi}
\email{gtlandi@if.usp.br}
\affiliation{Instituto de F\'isica da Universidade de S\~ao Paulo,  05314-970 S\~ao Paulo, Brazil}
\author{Mauro Paternostro}
\affiliation{Centre for Theoretical Atomic, Molecular and Optical Physics,
School of Mathematics and Physics, Queen's University Belfast, Belfast BT7 1NN, United Kingdom}

\begin{abstract}
Thermodynamic irreversibility is well characterized by the entropy production arising from  non-equilibrium quantum processes. We show that the entropy production of a quantum system undergoing open-system dynamics can be formally split into a  term that only depends on population unbalances, and one that is underpinned by quantum coherences. 
This allows us to identify a genuine quantum contribution to the entropy production in non-equilibrium quantum processes. 
We discuss how these features emerge both in Lindblad-Davies  differential maps and finite maps subject to the constraints of thermal operations. 
We also show how this separation naturally leads to two independent entropic conservation laws for the global system-environment dynamics, one referring to the redistribution of populations between system and environment  and the other describing how the coherence initially present in the system is distributed into local coherences in the environment and non-local coherences in the system-environment state.
Finally, we discuss how the processing of quantum coherences and the incompatibility of non-commuting measurements  leads to fundamental limitations in the description of quantum trajectories and fluctuation theorems. 
\end{abstract}
\maketitle{}

%%%%%%%%%%%%%%%%%%%%%%%%%%%%%%%%%%%%%%
%
%
%		INTRODUCTION
%
%
%%%%%%%%%%%%%%%%%%%%%%%%%%%%%%%%%%%%%%
\section*{\label{sec:int}Introduction}

Irreversible processes undergone by an open quantum system are associated with a production of entropy  that is fundamentally different from the possible flow of entropy resulting from the mutual coupling of the system with its  environment. 
%However, in addition, one may also observe an increase in entropy due to the irreversible dynamics. 
Such inevitable contribution to the entropy change of the state of a system is termed {\it entropy production}. If entropy is labelled as $S$, we describe its rate of change as 
\begin{equation}\label{dSdt}
\frac{\ud S}{\ud t} = \Pi - \Phi,
\end{equation}
where $\Phi$ is the entropy flux rate from the system to the environment and $\Pi$ is the entropy production rate. 
According to the second law of thermodynamics, we should have $\Pi \geq 0$, with $\Pi = 0$ if and only if the system is at equilibrium. 
The characterization of the degree of irreversibility of a process, and thus the understanding of entropy production is both fundamentally relevant and technologically desirable. On one hand, such understanding would provide the much needed foundations to the emergence of time-symmetry breaking entailed by irreversibility and epitomized, for instance, by seminal results such as Onsager's theory of irreversible currents \cite{Onsager1931,Onsager1931a,Machlup1953,DeGroot1961,Tisza1957}. On the other hand, a characterization of irreversible entropy could help us designing thermodynamically efficient quantum technologies~\cite{Escher2011,Verstraete2008a}. 

In general, the open dynamics of a quantum system gives rise to two processes. 
The first  corresponds to the transitions between energy levels of the system, which will cause the populations to adjust to the values imposed by the bath. 
Measures of the entropy production for this kind of processes have been known for many years, particularly in the context of Fokker-Planck equations \cite{Tome2010,Spinney2012,Landi2013b} and Pauli master equations \cite{Germany1976,Tome2012}.

The second process undergone by an open quantum system is the loss of coherence in the energy eigenbasis. 
Coherence is an essential resource for quantum processes~\cite{Streltsov2016}, likely representing the ultimate feature setting quantum and classical worlds apart. 
Only very recently have steps towards the formulation of a unified resource theory of coherence been made~\cite{Baumgratz2014}. 
While the role of quantum coherences in thermodynamics is yet to be fully understood (both qualitatively and quantitatively), it is known that coherence plays a role in the allowed transitions generated by thermal operations~%and which determine the second law of thermodynamics for microscopic systems 
\cite{Oppenheim2002,Horodecki2013,Lostaglio2015,Lostaglio2015a,Cwikli2015,Misra2016,Das2017}.
Moreover, it also affects the performance of non-adiabatic work protocols \cite{Korzekwa2016,Vacanti2015,Kammerlander2016,Francica2017}. 
%In fact, while finishing this paper we became aware of Ref.~\cite{Francica2017}, which in some aspects is complementary to the present paper. 

Understanding the interplay between population dynamics and loss of coherence represents a pressing problem in the field of thermodynamics of quantum systems. 
In particular, one is naturally led to wonder how entropy production is underpinned by the dynamics of quantum coherences in irreversible open dynamics. 

Shedding light on such an intimate relation is the main motivation of this paper, 
%While finishing this paper we became aware of Ref.~\cite{Francica2017}, where this problem is addressed in the context of unitary work protocols. 
where we put forward a formal description of entropy production in terms of two contributions,  one clearly related to the dynamics of populations and the other depending explicitly on the coherence within the state of the system undergoing the open process. % these processes within the context of quantum master equations and which is concerned with the entropy produced during the relaxation of a system toward thermal equilibrium. 
%Our work is therefore, in this sense, complementary to that of  Ref.~\cite{Francica2017}. 
We first discuss how these features emerge in the case of Davies-type master equations, for which the formulation is transparent. 
We then address more general dynamical maps satisfying the constraints of thermal operations \cite{Brandao2013}. 
In this case, we show how our result can be connected to recent resource theoretical developments~\cite{Lostaglio2015,Cwikli2015}. 

Afterwards, we address the main physical  implications of these two contributions to the entropy production. 
First, we discuss how it enables one to construct two independent entropic conservation laws for the global system-environment dynamics. 
The first is entirely classical and relates to the redistribution of populations between system and enviroment. 
The second, on the other hand, dictates  how the coherence initially prepared in the system is distributed among local coherences in the environment and  non-local coherences in the global system-environment state. 
Lastly, we address the issue of how to access entropy production by means of quantum measurements and  quantum trajectories. 
We show that even in the presence of coherence, it is possible to construct measurement protocols which satisfy fluctuation theorems. 
However, unlike in the case of closed systems \cite{Francica2017}, it is not possible to consider a single measurement protocol in which the fluctuation theorems are satisfied for both contributions of the entropy production individually.

Several advances in  the past decade have consistently shown that it is possible to engineer  systems in which thermodynamics  coexists with quantum effects. 
It is our hope that the framework put forth in this  paper may contribute for the development of a unified theory describing both thermal and quantum resources.

%%%%%%%%%%%%%%%%%%%%%%%%%%%%%%%%%%%%%%
%
%
%		Markovian systems
%
%
%%%%%%%%%%%%%%%%%%%%%%%%%%%%%%%%%%%%%%
\section*{Results}

\subsection*{\label{sec:ent} Entropy production due to the processing of coherence}

Coherence is a basis dependent concept and, in principle, no preferred basis exists. 
Here we adopt the viewpoint according to which a preferred basis only emerges when it is imposed by the environment~\cite{Zurek1981,Zurek2003b}, a perspective that is typically referred to as {\it einselection}.
There are several scenarios in which a preferred basis may emerge. We shall focus on two of them.
The first are Davies maps \cite{Alicki2008,Breuer2007,Gardiner2004}, which make use of the weak-coupling approximation, and the second are the so-called thermal operations \cite{Brandao2013,Lostaglio2015,Cwikli2015}. 
Both scenarios lead to energy conservation so that the energy eigenbases  are naturally selected as preferred bases. 
However, while in the weak-coupling scenario, energy conservation is taken as an approximation midway through the calculations, in the case of thermal operations this is imposed from the start, as an assumption of the model. 
Consequently, thermal operations are much more versatile, while simultaneously offering a much clearer physical interpretation.
We also assume, for simplicity, that the energy eigenvalues of the system are non-degenerate, since this would cause the preferred basis to depend on additional details of the system-environment interaction, which one seldom has access to.

\subsubsection*{Lindblad-Davies maps}

%For instance, consider a system with Hamiltonian $H_S$ such that $H_S |n\rangle = E_n^S |n\rangle$. 
%If the system weakly interacts with an environment that is in thermal equilibrium then it will eventually thermalize in the Gibbs state $\rho_S^\text{eq} = e^{-\beta H_S}/Z_S$, where $H_S$ is the system's Hamiltonian, $\beta$ is the inverse temperature of the bath and $Z_S = \tr(e^{-\beta H_S})$. 
%Consequently, in this case the  environment singled out the energy eigenbasis $|n\rangle$ as the preferred basis. 
%This is the type of physical scenario which will be considered in this paper.

We begin by analyzing the so-called Lindblad-Davies maps, which generally describe the weak contact of a system with a thermal environment. They have the form  \cite{Alicki2008,Breuer2007,Gardiner2004}
\begin{equation}\label{Lindblad}
\frac{\ud \rho_S}{\ud t} = - i [H_S,\rho_S] + D(\rho_S),
\end{equation}
where $\rho_S$ is the density matrix of the system, $H_S$ is the Hamiltonian  and $D(\rho_S)$ is a Lindblad dissipator having the Gibbs state $\rho^{\text{eq}}_S=e^{-\beta H_S}/Z_S$ as a fixed point; i.e. $D(\rho_S^\text{eq}) = 0$  (here $Z_S = \tr e^{-\beta H_S}$ is the partition function and $\beta$ the inverse temperature of the bath). 

Davies maps are known to lead  to a separation between the evolution of populations, $p_n = \langle n | \rho_S | n \rangle$, of the density matrix 
 and the  off-diagonal coherences $p_{nm} = \langle n | \rho_S | m \rangle$. Here, we have called $\{\vert{n}\rangle\}$ the elements of the basis imposed by the specific environment being considered. In the specific case of a Davies map, such basis is that of the energy eigenstates, whose set of eigenvalues we call $\{E_n\}$.

Following Eq.~\eqref{Lindblad}, the populations evolve following the Pauli master equation
\begin{equation}\label{Pauli}
\frac{\ud p_n}{\ud t} = \sum\limits_{m} \left[ W(n|m) p_m - W(m|n) p_n\right],
\end{equation}
where $W(n|m)$ are the transition rates from the energy level $E_n$ to level $E_m$. They satisfy the detailed-balance condition 
\begin{equation}\label{detailed_balance}
\frac{W(n|m)}{W(m|n)} = \frac{p_n^{\text{eq}}}{p_m^{\text{eq}}} = e^{-\beta (E_n - E_m)},
\end{equation}
where $p_n^{\text{eq}}=\langle n | \rho^{\text{eq}}_S | n \rangle$. As for the coherences, provided that the Bohr frequencies $\omega_{nm}=E_n-E_m$ are non-degenerate, they evolve independently of each other according to the equation
\begin{equation}
\frac{\ud p_{nm}}{\ud t} = - \bigg\{i \omega_{nm} -  \frac{1}{2} \sum\limits_k \left[ W(k|n) + W(k|m) \right]\bigg\} p_{nm}.
\end{equation}
As the second term in the right-hand side can be seen as an effective (generally temperature-dependent) damping term, the evolution of the coherences can be interpreted as that resulting from damped oscillations.

%\subsection{Non-equilibrium free energy and entropy production}

We now define  the non-equilibrium free energy as 
\begin{equation}
\label{Free}
F(\rho_S) =\tr(H_S \rho_S) +T \tr(\rho_S \ln \rho_S),
\end{equation} 
where $T$ is the temperature of the bath. While Eq.~(\ref{Free}) reduces to the usual expression $F_\text{eq} = - T \ln Z_S$ at equilibrium, for general non-equilibrium states, we can write
\begin{equation}\label{F1}
F(\rho_S) = F_\text{eq} + T S(\rho_S||\rho_S^\text{eq}),
\end{equation}
where  $S(\rho||\sigma) = \tr(\rho \ln \rho - \rho \ln \sigma)$ is the quantum relative entropy. As $S(\rho_S||\rho_S^\text{eq}) \geq 0$, we have that $F(\rho_S) \geq F_\text{eq}$. 
This condition thus defines the equilibrium state of a system as the state that minimizes the free energy~\cite{Callen1985}. 
Moreover, it establishes that, during relaxation, the free energy is a monotonically decreasing function of time whose value is determined by the distance, in state space, between the instantaneous state of the system $\rho_S$ and its equilibrium version $\rho_S^\text{eq}$.
Hence, one is naturally led to define the entropy production rate as 
\cite{Spohn1978,Breuer2003,Breuer2007,Deffner2011,Oliveira2016a}
\begin{equation}\label{Pi}
\Pi = - \frac{1}{T} \frac{\ud F(\rho_S)}{\ud t},
%= - \frac{\ud}{\ud t} S(\rho||\rho_\text{eq})
\end{equation}
which ensures that $\Pi \geq 0$ and  $\Pi = 0$ iff $\rho_S = \rho_S^\text{eq}$. 

Next we notice that  in terms of the eigenbasis of $H_S$, we can always split $S(\rho|| \rho_\text{eq})$ as 
\begin{equation}
S(\rho_S|| \rho_S^\text{eq}) = {\cal S}(p_S||p_S^\text{eq}) + {\cal C}(\rho_S).
\end{equation}
Here, ${\cal S}(p_S||p_S^\text{eq}) = \sum_n p_n \ln p_n/p_n^{\text{eq}}$ is the Kullback-Leibler divergence of the {\it classical} probability distribution entailed by the populations $p_S=\{p_n\}$ from that of the equilibrium state $p_S^\text{eq}=\{p_n^\text{eq}\}$. Moreover, we have introduced the relative entropy of coherence~\cite{Baumgratz2014}
\begin{equation}\label{C}
{\cal C}(\rho_S) = S(\Delta_{H_S}(\rho_S))  - S(\rho_S),
\end{equation}
where $\Delta_{H_S}(\rho)$ is the dephasing map, acting on the density matrix $\rho_S$, which removes all coherences from the various energy eigenspaces of $H_S$. %We shall loosely refer to this as a dephasing operation.
With this at hand, Eq.~(\ref{F1}) becomes
\begin{equation}\label{F2}
F(\rho_S) = F_\text{eq} + T {\cal S}(p_S||p_S^\text{eq}) + T {\cal C}(\rho_S).
\end{equation}
This is one of the central results of this work: It shows that \emph{quantum coherence is actually a part of the non-equilibrium free energy}, and thus contributes significantly to the determination of the non-equilibirum thermodynamics stemming from Eq.~\eqref{Lindblad}. The second term in $F(\rho_S)$ quantifies the increase in free energy due to population imbalance with respect to the equilibrium configuration and, as such, is a purely classical term. The last term, which is of a genuine quantum nature, determines the surplus in free energy that a non-equilibrium state with quantum coherences offers with respect to its diagonal (and thus classical) counterpart. 
\noindent

Let us now use the formal splitting in Eq.~(\ref{F2}) to recast the entropy production rate in Eq.~(\ref{Pi}) as
\begin{equation}\label{Pi_sep}
\Pi = \Pi_d + \Upsilon.
\end{equation}
The first term is written as 
\begin{equation}
\label{Pid}
\begin{aligned}
\Pi_d &= - \frac{\ud }{\ud t} {\cal S}(p_S|| p_S^{\text{eq}})\\
&=
\frac{1}{2} \sum\limits_{n,m} \left[ W(n|m) p_m - W(m|n) p_n \right] \ln \frac{W(n|m) p_m}{W(m|n) p_n}.
\end{aligned}
\end{equation}
and is then exactly the classical result derived in~\cite{Germany1976,Tome2012}.
The second contribution reads 
\begin{equation}
\label{cohe}
\Upsilon = - \frac{\ud {\cal C}(\rho_S)}{\ud t},
\end{equation}
which thus shows that the rate of loss of coherence that might ensue from the dynamics of the system enters quantitatively into the entropy production: the rate at which entropy is produced in a process where quantum coherences are destroyed as a result of the coupling with an environment surpasses the corresponding classical version. %This therefore provides the physical interpretation to the mismatch between entropy production arising from quantum and classica.
Clearly both $\Pi_d$ and $\Upsilon$ are non-negative and null only for $\rho_S = \rho_S^\text{eq}$. 
%Thus, we see that the classical contribution is augmented with an extra quantum effect due to the loss of coherences. 

Finally, let us address the entropy flux defined in  Eq.~(\ref{dSdt}). Using Eq.~(\ref{Pi}) we find 
%\begin{equation}
%\Phi = %\Pi - \frac{\ud S}{\ud t} = 
%\tr \left[ \frac{\ud \rho_S}{\ud t} \ln \rho_S^\text{eq}\right] = \sum\limits_n \frac{\ud p_n}{\ud t} \ln p_n^{\text{eq}}.
%\end{equation}
\begin{equation}\label{Phi_PhiE}
\Phi = \frac{\Phi_E}{T} = - \frac{1}{T} \sum\limits_n E_n^S \frac{\ud p_n}{\ud t},
\end{equation}
where $\Phi_E$ is the energy flux from the system to the environment. In deriving this equation we have used $p_n^{\text{eq}} = \exp[- \beta E_n^S]/Z_S$.
The entropy flux has thus no contribution arising from quantum coherences: entropy (and energy) will only flow due to imbalances in the populations. 
Any loss of coherence contributes only to the entropy production rate and has no associated flux.
It is important to emphasize that this is a feature of the present type of master equation. 
In other situations, such as strong coupling dynamics, the coherences in the system may play a relevant role in the entropy and heat fluxes.

%In summary, we see that when quantum systems are allowed to relax in contact with thermal reservoirs, there will be a clear separation between classical and quantum processes. The former involve only changes in populations and is responsible for both an entropy flux to the environment and an irreversible entropy production. 
%The quantum part, on the other hand, has no associated flux, but only an entropy production related to the inevitable loss of coherence. 

%%%%%%%%%%%%%%%%%%%%%%%%%%%%%%%%%%%%%%
%
%
%		Thermal operations
%
%
%%%%%%%%%%%%%%%%%%%%%%%%%%%%%%%%%%%%%%
%\section{\label{sec:therm}Non-Markovian systems and thermal operations}

\subsubsection*{\label{sec:therm} Thermal Operations}

We now address the case of more general maps with the aim of gaining access to the environmental degrees of freedom, hence enhancing our understanding of the two contributions to the entropy production from the perspective of the joint system-environment properties.

We consider explicitly thermal operations, which have been the subject of numerous recent investigations in the context of resource theories \cite{Brandao2013,Lostaglio2015,Cwikli2015}. 
A thermal operation is physically described as the interaction of the system with an arbitrary environment, initially prepared in equilibrium $\rho_E^\text{eq} = e^{-\beta H_E}/Z_E$, through a unitary $U$ which conserves the total energy, that is an operation such that $[U, H_S+H_E]=0$. 
In this sense, the thermal operation hypothesis reminds us of the framework set by the weak-coupling approximation. 
However, it allows us to go significantly beyond the limitations of  weak-coupling, and thus extend our analysis to a larger set of physically meaningful cases \cite{Halpern2015a}.
We also call attention to the fact that in thermal operations the energy conservation condition is only imposed on the global unitary $U$, irrespective  of how this is generated. 
One way to do so is by means of a time-dependent interaction that is turned on and off. 
Another, much simpler approach, is to simply have any potential $V$ which satisfies $[V, H_S+H_E] = 0$. This will then generate a time-independent unitary $U = e^{- (H_S+H_E+V)t}$ which will be energy conserving.
This type of thermal operation was recently used in Ref.~\cite{Micadei2017} to study the heat exchange between two qubits in a magnetic resonance setup.

The state of the composite system after the evolution in a thermal operation will be 
\begin{equation}\label{map_SE}
\rho_{SE}' = U (\rho_S \otimes \rho_E^\text{eq}) U^\dagger.
\end{equation}
We label the environmental energies and eigenstates as $\{E_{\mu}^E\}$ and $\{|\mu\rangle\}$, respectively. We also call
$q_\mu = e^{-\beta E_\mu^E}/Z_E$  its initial thermal populations. 
Energy conservation then implies that 
\begin{equation}\label{energy_conservation}
\langle m,\nu| U |n,\mu\rangle \propto \delta(E_n^S + E_\mu^E - E_m^S - E_\nu^E).
\end{equation}
Tracing out the environment one obtains the Kraus map  for the system
\begin{equation}\label{Thermal_Kraus}
\rho_S' = \tr_{E} \bigg[ U(\rho_S \otimes \rho_E^\text{eq}) U^\dagger\bigg] = \sum\limits_{\mu,\nu} M_{\mu,\nu} \rho_S M_{\mu,\nu}^\dagger
\end{equation}
where $M_{\mu,\nu} = \sqrt{q_\mu} \langle \nu | U | \mu \rangle$. 
Clearly, the Gibbs state $\rho_S^\text{eq}$ is a fixed point of this equation. 
Moreover, the Davies maps studied in the previous Section correspond to particular Markovian limits of Eq.~\eqref{Thermal_Kraus}. 
%One could also trace out the system and obtain the corresponding quantum operation for the environment. 
%In this case the final expression will be slightly more cumbersome. 
%We note however, that if the eigenvalues of $H_S$ are non-degenerate, it follows that 

The energy conservation condition implies that, when the eigenvalues of $H_S$ are non-degenerate, Eq.~(\ref{Thermal_Kraus}) is an incoherent operation in the sense of Ref.~\cite{Baumgratz2014}.
That is, defining the  energy eigenstates as the set of incoherent states,  this process always maps incoherent states into incoherent states.
This in turn allows for an independent processing of both populations and coherences. 
The diagonal entries will, in particular, evolve according to the classical Markov chain 
\begin{equation}\label{Markov}
p_m' = \sum\limits_n Q(m|n) p_n,
\end{equation}
where $Q(m|n) = \sum_{\mu,\nu} |\langle m |M_{\mu,\nu}| n \rangle |^2$ is the transition probability from state $n$ to state $m$, a quantity playing the role of 
the transition rate $W(m|n)$ in Eq.~(\ref{Pauli}). The processing of the coherences, on the other hand, takes place independently of the changes in populations. 
In particular, if the Bohr frequencies $\omega_{mn}$ are non-degenerate, this processing takes the simple form 
\begin{equation}
p_{n,m}' = \alpha_{n,m} p_{n,m},
\qquad
\alpha_{n,m} = \sum_{\mu,\nu} \langle n | M_{\mu,\nu} | n \rangle \langle m |M_{\mu,\nu}^\dagger |m\rangle.
\end{equation} 
As shown in Ref.~\cite{Cwikli2015}, the processing of coherence is not independent of the population changes, but must satisfy 
the inequality
\begin{equation}
|\alpha_{n,m}|^2 \leq Q(n|n) Q(m|m),
\end{equation}
which provides a bound to the maximum amount of coherence that may be lost in a thermal operation.

We now turn to the analysis of the entropy production in this scenario. 
Unlike the previous Section, as the dynamics in Eq.~\eqref{Thermal_Kraus} is in general non-Markovian and we only have access to the global map, it is not possible to address the rate of entropy production $\Pi$, but only the total entropy $\Sigma$ produced in the process. 
In this case, using the contractive property of the relative entropy~\cite{Lindblad1975}, we have 
$S(\rho_S'|| \rho_S^\text{eq}) \leq S(\rho_S|| \rho_S^\text{eq})$.
Consequently, the free energy  Eq.~(\ref{F1}) remains a non-increasing function,
thus justifying the following definition of total entropy production
\begin{equation}\label{Sigma}
\Sigma = - \frac{\Delta F}{T}  =  S(\rho_S|| \rho_S^\text{eq}) - S(\rho_S'|| \rho_S^\text{eq}) \geq 0.
\end{equation}
This expression may be taken as a basic postulate in our framework, motivated by physical consistency arguments that can be even reinforced by a quantum trajectory point of view, as discussed e.g. in Refs.~\cite{Manzano2017a,Manzano2015}. Other approaches have also been used elsewhere~\cite{Esposito2010a,Brandao2015}.

As $\Sigma$ is also based on the quantum relative entropy, a splitting akin to Eq.~(\ref{Pi_sep}) into non-negative population-related and a coherence-related terms is in order, and we have 
\begin{equation}\label{Sigma_sep}
\Sigma = \Sigma_d + \Xi,
\end{equation}
where 
\begin{IEEEeqnarray}{rCl}
\label{Sigma_d}
\Sigma_d &=& S(p_S||p_S^\text{eq})- S(p_S'||p_S^\text{eq}),\\[0.2cm]
\label{Theta}
\Xi &=&  {\cal C}(\rho_S) - {\cal C}(\rho_S').
\end{IEEEeqnarray}
%These quantities are always non-negative. 
The non-negativity of $\Sigma_d$ follows immediately from the fact that in thermal operations diagonal elements evolve independently of coherences. 
%Hence, $\Sigma_d\geq 0$ for the exact same reason that $\Sigma\geq0$.
The positivity of~~$\Xi$, on the other hand, follows from the fact that a thermal operation is incoherent~\cite{Baumgratz2014}. 
In the limit where the Davies maps are recovered,  $\Sigma_d$ and $\Xi$ become respectively the integrated versions of $\Pi_d$ and $\Upsilon$ in Eq.~(\ref{Pi_sep}).
This demonstrates the generality, under suitable and reasonable assumptions on the form of the system-environment coupling, of the central result of our investigation.

%%%%%%%%%%%%%%%%%%%%%%%%%%%%%%%%%%%%%%
%
%
\subsection*{Implications of the central results}
%
%
%%%%%%%%%%%%%%%%%%%%%%%%%%%%%%%%%%%%%%

We now explore which considerations  can be drawn in light of the formal splitting of the entropy production demonstrated above.

%%%%%%%%%%%%%%%%%%%%%%%%%%%%%%%%%%%%%%
%
%
\subsubsection*{\label{sec:cons}Entropic conservation laws}
%
%
%%%%%%%%%%%%%%%%%%%%%%%%%%%%%%%%%%%%%%

The structure of thermal operations  and Eq.~(\ref{map_SE}) imply a series of conservation rules for the processing of populations and coherences. 
First, energy conservation implies that the total entropy production in Eq.~(\ref{Sigma}) can be written as~\cite{Esposito2010a,Strasberg2016}
\begin{equation}\label{conservation_Sigma_2}
\Sigma = S(\rho_E'||\rho_E^\text{eq}) + \mathcal{I}(\rho_{SE}'),
\end{equation}
where $\rho_E'=\tr_S(\rho'_{SE})$  and $\mathcal{I}(\rho_{AB}) = S(\rho_A) + S(\rho_B) - S(\rho_{AB})$ is the mutual information of a bipartite system. 
This gives an interesting interpretation of $\Sigma$ as being related to the change in the environmental state, measured by the first term, and the total degree of correlations created by the thermal operation, measured by the mutual information.
As discussed in Ref.~\cite{Strasberg2016}, Eq.~(\ref{conservation_Sigma_2}) also provides a clear interpretation of how irreversibility emerges from a global unitary dynamics, ascribing it to two reasons. One is the creation of correlations between system and environment, which are never recovered once one traces out the environment (hence  giving rise to an irretrievable loss of information). 
The second is related to the fact that the system pushes the environment away from equilibrium. 
As shown in Ref.~\cite{Manzano2017a}, if $\rho_E' - \rho_E^\text{eq} \sim \epsilon$ then $S(\rho_E'|| \rho_E^\text{eq}) \sim \epsilon^2$, whereas $S(\rho_E') - S(\rho_E^\text{eq}) \sim \epsilon$. 
Thus, when the reservoir is large, the first term becomes negligible and the main contribution to the entropy production will come from the total correlations created between system and environment.

Next we note that as the map~(\ref{map_SE}) is unitary, it follows that $S(\rho_{SE}') = S(\rho_{SE})$. 
However, as $[U,H_S+H_E]=0$, 
the dephasing operation $\Delta_{H_S+H_E}$ commutes with the unitary evolution so that, in addition to the total entropy being conserved, the same is also true for the dephased entropies 
\begin{IEEEeqnarray}{rCl}
\label{diagonal_conservation}
S(\Delta_{H_S+H_E}(\rho_{SE}')) 
%&=& S(\Delta_{H_S+H_E}(\rho_{SE})) \\[0.2cm]
&=& S(\Delta_{H_S}(\rho_S)) + S(\rho_E^\text{eq}).
\end{IEEEeqnarray}
This result reflects how the changes in population in the system and environment affect the  information content of the diagonal elements of $\rho_{SE}'$. 
Note that the left-hand side contemplates, at most, coherences in the degenerate subspaces of $H_S+H_E$, which are not resources from the perspective of this operation. 

From Eqs.~(\ref{conservation_Sigma_2}) and ~(\ref{diagonal_conservation}), it follows that a similar law must also hold for the relative entropy of coherence
\begin{equation}\label{coherence_conservation}
{\cal C}(\rho_{SE}') = {\cal C}(\rho_S).
\end{equation}
Where ${\cal C}(\rho_{SE}')=S(\Delta_{H_S+H_E}(\rho_{SE}'))-S(\rho_{SE}')$.
Thus, we see that the reduction in the coherence of the system after the map is due to a redistribution of this coherence over the global system-environment state. 

Substituting Eq.~(\ref{coherence_conservation}) into Eq.~(\ref{Theta}) shows that the contribution of the entropy production due to quantum coherences may be written as 
\begin{equation} \label{eq:xi}
\Xi = {\cal C}(\rho_{SE}') - {\cal C}(\rho_S').
\end{equation}
Thus, the entropy production due to quantum coherences can be seen as the mismatch between the global coherences in the correlated system-environment state and the local coherences in the final state.
We can also relate $\Xi$ to the notion of correlated coherence, introduced recently in~\cite{Tan2016} and defined as
\begin{equation}\label{eq:cc}
{\cal C}_{cc}(\rho_{SE}') = {\cal C}(\rho_{SE}') - {\cal C}(\rho_{S}') - {\cal C}(\rho_{E}') \geq 0,
\end{equation}
where ${\cal C}(\rho_{E}')=S(\Delta_{H_E}(\rho_E'))-S(\rho_E')$.
This quantity therefore represents the portion of coherence that is contained within the correlations between system and environment. 
Combining the Eqs.~(\ref{eq:xi}) and (\ref{eq:cc}), it is then possible to write
\begin{equation}\label{eq:Xicc}
\Xi = {\cal C}(\rho'_E) +  {\cal C}_{cc}(\rho'_{SE}).
\end{equation}
This has the same form as Eq.~(\ref{conservation_Sigma_2}), with the first term representing the local coherences developed in the environment and the second term representing the non-local contribution. 
Thus, we may conclude from this result that entropy production has a clearly local contribution, related to the creation of coherences in the environment, and a non-local contribution related to the creation of shared coherences in the system-environment state.

%%%%%%%%%%%%%%%%%%%%%%%%%%%%%%%%%%%%%%
%
%
\subsubsection*{\label{sec:fluc}Stochastic trajectories and fluctuation theorems}
%
%
%%%%%%%%%%%%%%%%%%%%%%%%%%%%%%%%%%%%%%
%{\bf I think this can go to an appendix}
Lastly, let us consider the stochastic version of the entropy production arising from quantum trajectories. 
In order to correctly treat the coherences present in the system, we adopt the following procedure.
In the forward protocol, the environment is prepared in the thermal state $\rho_E = \sum_\mu q_\mu |\mu\rangle\langle \mu|$, where 
$q_\mu = e^{-\beta E_\mu^E}/Z_E$.
The system, on the other hand, is taken to be in an arbitrary state $\rho_S = \sum_\alpha p_\alpha |\psi_\alpha \rangle \langle \psi_\alpha|$, which in general contain coherences, so that the basis $|\psi_\alpha\rangle$ is  incompatible with the energy basis $|n\rangle$. 
As the first step in the protocol, we then perform local measurements in the bases $|\psi_\alpha\rangle$ and $|\mu\rangle$ of S and E, obtaining the state $|\psi_\alpha,\mu\rangle$  with probability $p_\alpha q_\mu$. 
Next we evolve both with a joint unitary $U$.
Finally, in the third step we measure only the environment, again in the energy basis $|\nu\rangle$.
Due to the measurement backaction, the state of the system then collapses to the pure state
\begin{equation}\label{Phi_F}
|\Phi_{\text{F}|\alpha\mu \nu}\rangle = \frac{\langle \nu | U | \psi_\alpha, \mu\rangle}{\sqrt{P_F(\nu|\alpha\mu)}},
\end{equation}
where $P_F(\nu|\alpha\mu) = ||\langle \nu | U | \psi_\alpha, \mu\rangle||^2$. 
For a discussion on the effects of choosing different bases for the second measurement in the environment, see Ref.~\cite{Manzano2017a}.

The final state  $\rho_S'$ of the system will then be given by an ensemble average over all possible final states~(\ref{Phi_F}), weighted by the probability of the stochastic trajectory $(\alpha, \mu, \nu)$; viz., 
\begin{equation}\label{rhoSp_trajectories}
\rho_S' = \sum\limits_{\alpha, \mu, \nu} P_F(\nu|\alpha,\mu) p_\alpha q_\mu |\Phi_{\text{F}|\alpha\mu \nu}\rangle \langle \Phi_{\text{F}|\alpha\mu \nu}|.
\end{equation}
One may directly verify that this state is indeed equal to the unmeasured final state $\rho_S'$ in Eq.~(\ref{Thermal_Kraus}).
The states~(\ref{Phi_F}), however, are not the eigenstates of $\rho_S'$ and in fact  don't even form a basis. The diagonal structure of $\rho_S'$ will thus be of the form $\rho_S' = \sum_{\beta} p_\beta' |\psi_\beta'\rangle\langle \psi_\beta'|$, where $|\psi_\beta'\rangle$ is a new basis set that is not trivially related to neither $|\psi_\alpha\rangle$ nor $|\Phi_{\text{F}|\alpha\mu \nu}\rangle$. 
This is the key difference that appears due to the presence of coherences (if the initial state of the system were diagonal, the same would be true for the final state, since this is a thermal operation). 
The probabilities   $p_\beta'$ will be given by 
\begin{equation}
p_\beta' = \sum\limits_{\alpha, \mu, \nu} p_{\beta|\alpha, \mu, \nu} \; P_F(\nu|\alpha, \mu) \; p_\alpha \; q_\mu \; ,
\end{equation}
where $p_{\beta|\alpha, \mu, \nu} = |\langle \psi_\beta' | \Phi_{\text{F}|\alpha\mu\nu}\rangle|^2$ is the conditional probability of finding the system in $|\psi_\beta'\rangle$ given that it ended the forward protocol in $|\Phi_{\text{F}|\alpha\mu \nu}\rangle$.

The stochastic trajectory generated by the measurement outcomes is specified by the three quantum numbers $(\alpha, \mu, \nu)$. 
However, following Ref.~\cite{Park2017}, we may augment the trajectory (an idea first introduced by Dirac \cite{Dirac1945}) by introducing $\beta$ as an additional quantum number, so that the trajectory $\mathcal{X} = (\alpha, \mu, \beta, \nu)$ is defined by the probability 
\begin{equation}\label{path_prob_F}
\mathcal{P}_F[\mathcal{X}] = p_{\beta|\alpha, \mu, \nu}\; P_F(\nu|\alpha, \mu) \; p_\alpha \; q_\mu .
\end{equation}
Indeed, using the definition~(\ref{Phi_F}),   one readily finds that
\begin{equation}\label{path_simplify}
p_{\beta|\alpha, \mu, \nu}\; P_F(\nu|\alpha, \mu)  = |\langle \psi_\beta', \nu | U | \psi_\alpha, \mu \rangle |^2 = P(\beta,\nu | \alpha, \mu),
\end{equation}
which is nothing but the transition probability of observing a transition from $|\psi_\alpha, \mu\rangle$ to $|\psi_\beta', \nu\rangle$.

Next we define the backward protocol. 
The initial state of the system is drawn from one of the possible  eigenstates $|\psi_\beta'\rangle$ of $\rho_S'$ \cite{Elouard2017a}, whereas the environment is taken to be in equilibrium and is again measured in the energy basis $|\nu\rangle$.
This yields the state $|\psi_\beta', \nu\rangle$ with probability $p_\beta' q_\nu$. 
We then apply the time-reversed unitary $U^\dagger$ and, in the end, measure $E$ in the basis $|\mu\rangle$. 
As a consequence the system collapses to 
\begin{equation}\label{Phi_B}
|\Phi_{\text{B}|\beta\nu \mu}\rangle = \frac{\langle \mu | U^\dagger | \psi_\beta', \nu\rangle}{\sqrt{P_B(\mu|\beta,\nu)}},
\end{equation}
where $P_B(\mu|\beta,\nu) = ||\langle \mu | U^\dagger | \psi_\beta', \nu\rangle||^2$.

The backward trajectory is specified by the quantum numbers $(\beta, \nu, \mu)$. 
However, as in the forward case, we can define the augmented trajectory $\mathcal{X} = (\alpha, \mu, \beta, \nu)$ by introducing the conditional probability $p_{\alpha|\beta, \nu,\mu} = |\langle \psi_\alpha | \Phi_{\text{B}|\beta\nu\mu}\rangle|^2$. 
The probability for the augmented backward trajectory will then be given by 
\begin{equation}\label{path_prob_B}
\mathcal{P}_B[\mathcal{X}] = p_{\alpha|\beta, \nu,\mu}\; P_B(\mu|\beta,\nu) \; p_\beta' \; q_\nu .
\end{equation}
With the path probabilities~(\ref{path_prob_F}) and (\ref{path_prob_B}), we can now define the entropy production in the usual way, as 
\begin{equation}\label{sigma_traj_def}
\sigma[\mathcal{X}] = \ln \frac{\mathcal{P}_F[\mathcal{X}] }{\mathcal{P}_B[\mathcal{X}] }.
\end{equation}
By construction, $\sigma$ satisfies a detailed fluctuation theorem \cite{Crooks1998,Manzano2017a,Elouard2017a}
\begin{equation}\label{FT}
\langle e^{-\sigma[\mathcal{X}] } \rangle = 1.
\end{equation}
Moreover, similarly to Eq.~(\ref{path_simplify}), it follows that $p_{\alpha|\beta, \nu,\mu}\; P_B(\mu|\beta,\nu) = P(\beta,\nu|\alpha, \mu)$, so that 
\begin{equation}\label{sigma_traj}
\sigma[\mathcal{X}] = \ln \frac{p_{\beta|\alpha, \mu, \nu}\; P_F(\nu|\alpha, \mu) \; p_\alpha \; q_\mu }{p_{\alpha|\beta, \nu,\mu}\; P_B(\mu|\beta,\nu) \; p_\beta' \; q_\nu} = \ln \frac{p_\alpha q_\mu}{p_\beta' q_\nu}.
\end{equation}
Thus, we see that the conditional terms cancel out. 
Physically, this means that there is no additional entropic cost in introducing the augmented trajectories, which is a consequence of the fact that the augmentation was done using the eigenstates $|\psi_\beta'\rangle$ of $\rho_S'$. 
One may also directly verify that  $\langle \sigma[\mathcal{X}]  \rangle = \Sigma$ is the average entropy production in Eq.~(\ref{Sigma}).

Next we address the question of how to define stochastic quantities for the two contributions to the entropy production in  Eq.~(\ref{Sigma_sep}).
That is, we wish to separate 
\begin{equation}\label{stoch_separation}
\sigma[\mathcal{X}] = \sigma_d[\mathcal{X}] + \xi[\mathcal{X}],
\end{equation}
such that $\langle \sigma_d[\mathcal{X}]  \rangle = \Sigma_d$ and $\langle \xi[\mathcal{X}]  \rangle = \Xi$.
This can be accomplished by augmenting the trajectory once again to include the populations of the system in the energy basis. That is, we define $\tilde{\mathcal{X}} = \{\alpha, \mu, n, \beta, \nu, m\}$, 
with associated path probabilities
$\mathcal{P}_{F(B)}[\tilde{\mathcal{X}}] = \mathcal{P}_{F(B)}[\mathcal{X}] p_{n|\alpha} p'_{m|\beta}$, where 
we defined the  conditional probabilities $p_{n|\alpha} = |\langle n |\psi_\alpha \rangle|^2$ and $p_{m|\beta} = |\langle m | \psi_\beta'\rangle|^2$.

%Of course, these definitions are of limit extend, due to the incompatibility imposed by the quantum rules in simultaneously knowing the state in multiple bases. 
%They therefore represent the maximum amount of information given this incompatibility. 

We then define the stochastic quantities
\begin{IEEEeqnarray}{rCl}
\label{sigma_d}
\sigma_d[\tilde{\mathcal{X}}] &=&  \ln \bigg( \frac{p_n q_\mu}{p_m' q_\nu}\bigg),\\[0.2cm]
\label{xi}
\xi[\tilde{\mathcal{X}}] &=&  \ln \bigg( \frac{p_\alpha p_m'}{p_\beta' p_n}\bigg),
\end{IEEEeqnarray}
where $p_n = \langle n | \rho_S | n \rangle$ and $p_m' = \langle m |\rho_S' |m \rangle$ are the populations in the energy eigenbasis at the initial and final states [cf. Eq.~(\ref{Markov})]. 
Summing  these two contributions immediately yields Eq.~(\ref{sigma_traj}).
Moreover, one may also verify that $\langle \sigma_d[\tilde{\mathcal{X}}]  \rangle = \Sigma_d$ and $\langle \xi[\tilde{\mathcal{X}}]  \rangle = \Xi$.
Hence, these quantities do indeed represent the stochastic counterparts of the two contributions to the entropy production.

A second glance at Eq.~(\ref{xi}) reveals that, on the stochastic level, the entropy production $\xi$ due to the loss of coherence,  is nothing but the change in  information between  incompatible bases, a quantity sometimes referred to as \emph{information gain} \cite{Groenewold1971}.
Hence, we arrive at the important conclusion that the incompatibility between classical and quantum entropy production can be traced back, at the stochastic level, to the basis incompatibility of the quantum rules.
Indeed, if we  rewrite the  fluctuation theorem~(\ref{FT}) as 
$\langle   e^{-\sigma_d[\tilde{\mathcal{X}}]  - \xi[\tilde{\mathcal{X}]}} \rangle = 1$,
we can clearly see that due to the presence of coherences, the classical fluctuation theorem, that one finds for diagonal initial states, is not satisfied. 
Instead, it must be corrected by the information gain. 
It is also possible to draw an alternative interpretation in terms of the entropy production due to quantum measurements, as studied for instance in Refs.~\cite{Manzano2015,Elouard2017a}. 
When a measurement is performed in a basis which commutes with the system's density matrix, no entropy is produced. 
Non-commuting measurements, on the other hand, have an associated entropy production related to the loss of coherence. 
This is precisely the content of Eq.~(\ref{xi}).
No such additional entropy production was generated for the first augmentation that led us to Eq.~(\ref{path_prob_F}), as in this case there is no information gain since the basis $|\psi_\beta'\rangle$ is the basis that diagonalizes $\rho_S'$.

\section*{Discussion}

We have addressed the role played by quantum coherence in determining the behaviour of the entropy production, a fundamental quantifier of thermodynamic irreversibility. By making physically reasonable assumptions on the form of the dynamics undergone by a quantum system and its environment, we have been able to single out the contribution that quantum coherences, a genuine non-classical feature of the state of a given dynamical system, play in quantifying the rate of irreversible entropy production. Such contribution appears to be fully distinct from the one arising from unbalances in the energy eigenbasis of the state of the system, which brings about a classical flavour. Moreover, it can be interpreted in a physically transparent manner as the {\it thermodynamic cost} that one has to pay for the destruction of coherences that were seeded in the state of the system itself. In turn, our results have interesting consequences for the interpretation of the process of producing entropy as a result of the dynamical generation of correlations (or equivalently coherences) between a quantum system and its environment.

We believe that the theory put forth in this paper may prove  a useful step forward towards the setting up of a self-contained framework for the interpretation of thermodynamic irreversibility at the quantum nanoscale, which is still sorely missing despite the key role that entropy production will play in the quantification of the thermodynamic fingerprint of managing quantum dynamics. 
For instance, it could serve as a starting point for the development of a theory of quantum entropy production in non-equilibrium steady-states of systems connected to multiple reservoirs. Or, as a tool for quantifying the contribution of loss of coherence in the operation of finite-time quantum heat engines.

\section*{Data availability statement}

No data sets were generated or analyzed during the current study.

\section*{Acknowledgments} 

GTL acknowledges the funding from the S\~ao Paulo Research agency, under grant number 2016/08721-7. MP acknowledges funding from the SFI-DfE Investigator Programme, the EU H2020 Collaborative Project TEQ, the Leverhulme Trust, and the Royal Society. 
JPS and GTL acknowledge the Brazilian funding agency CAPES. 

\section*{Additional Information}

{\bf Competing interests:} The authors declare no competing interests.

\section*{Author contributions}

JPS developed the detailed formal calculations in collaboration with GTL; LCC, GTL, and MP conceived the original idea and shaped it with the significant help of JPS. 
All authors contributed to the research and the preparation of the manuscript. 

%\bibliographystyle{naturemag}

%

%\bibliography{/Users/gtlandi/Documents/library}
%\bibliography{library}

\end{document}